\documentclass[prl,twocolumn,superscriptaddress,floatfix,showpacs]{revtex4}
\usepackage{graphicx}
\usepackage{bm}
\usepackage{amsmath}
\usepackage{amsfonts}
\usepackage{amssymb}
\usepackage[thinspace,squaren,pstricks,derivedinbase,derived]{SIunits}

\renewcommand{\vec}[1]{{\mathbf #1}}

\begin{document}
\title{Asymmetric Spin-wave Dispersion on Fe(110): Direct Evidence of Dzyaloshinskii--Moriya Interaction}

\author{Kh.~Zakeri}
\email[Corresponding author: ]{zakeri@mpi-halle.de}
\author{Y.~Zhang}
\author{J.~Prokop}
\author{T.-H. Chuang}
\author{N.~Sakr}
\affiliation{Max-Planck-Institut f\"ur Mikrostrukturphysik,
Weinberg 2, 06120 Halle, Germany}
\author{W.~X.~Tang}

\affiliation{Max-Planck-Institut f\"ur Mikrostrukturphysik, Weinberg 2, 06120 Halle, Germany}
\affiliation{School of Physics, Monash University, Victoria 3800, Australia}
\author{J.~Kirschner}
\affiliation{Max-Planck-Institut f\"ur Mikrostrukturphysik, Weinberg 2, 06120 Halle, Germany}

\date{\today}
\begin{abstract}
The influence of the Dzyaloshinskii--Moriya interaction on the spin-wave dispersion in an Fe double-layer grown on W(110) is measured for the first time. It is demonstrated that the Dzyaloshinskii--Moriya interaction breaks the degeneracy of spin-waves and leads to an asymmetric spin-wave dispersion relation. An extended Heisenberg spin Hamiltonian is employed to obtain the longitudinal component of the  Dzyaloshinskii--Moriya vectors from the experimentally measured energy asymmetry.
\end{abstract}
\pacs{75.30.Ds,75.70.Ak,75.70.Rf,75.50.Bb}
\maketitle

In 1957, Dzyaloshinskii proposed an antisymmetric exchange interaction, based on symmetry arguments, to explain the weak ferromagnetism observed in some oxide materials e.g. $\alpha-$Fe$_2$O$_3$ (Hematite) \cite{Dzyaloshinskii}. Only three years later it was shown by Moriya  that, in principle, this interaction can be analytically derived by considering the relativistic spin-orbit correction in the Hamiltonian \cite{Moriya}. The antisymmetric exchange interaction, Dzyaloshinskii--Moriya (DM) interaction, became very important to understand many physical properties of different systems i.e. spin-glasses \cite{Fert}, cuprates \cite{Coffey}, molecular magnets \cite{Zhao,Raedt} and multiferroics \cite{Sergienko,Hu}.

In nanomagnetism, where the surface effects are noticeable, the spin-orbit coupling is one of the most important intrinsic magnetic perturbations, which creates novel phenomena. Recently, it has been shown that a strong spin-orbit coupling in the presence of the broken inversion symmetry at the surface leads to the DM interaction, which stabilizes a noncollinear spin structure for a Mn monolayer on W(110) \cite{Bode07} and W(100) \cite{Ferriani08} surfaces.

An ultrathin Fe film grown on W(110) is another system that is believed to show the DM interaction \cite{Vedmedenko,Heide,Meckler}.
Magnetic excitations in this quasi-two-dimensional spin system have been studied theoretically since many years \cite{Bloch30,Yafet86,Bruno91,Kambersky99,Muniz02,Udvardi03,Pini05,Costa08}. In the description of the collective magnetic excitations, only the symmetric exchange interaction was considered and the DM interaction has been neglected. In such systems, where DM interaction is relatively large, it should, in principle, change the intrinsic properties of the spin-waves (SWs). Only very recently, the influence of the DM interaction on the spin-wave dispersion has been predicted to give rise to an asymmetric spin-wave dispersion in an Fe monolayer on W(110) \cite{Udvardi09}. However, the effect of the DM interaction on the spin-wave dispersion in low dimensional systems has never been measured experimentally.

In this Letter we report the first experimental evidence of the influence of DM interaction on the spin-wave dispersion in a double-layer Fe. We show that in the presence of the DM interaction the spin-wave dispersion is asymmetric. By measuring the highly resolved spin polarized electron energy loss (SPEEL) spectra in both energy loss and gain regimes and by reversing the magnetization of the film, we measure the DM interaction driven asymmetry in the spin-wave dispersion of Fe double-layer grown on W(110). By using an extended Heisenberg spin Hamiltonian, the energy asymmetry is modeled giving rise to a quantitative determination of the longitudinal components of DM vectors.

The iron layer was deposited onto a clean W(110) single crystal at room temperature (RT). Special care has been taken concerning the cleaning of the W crystal as described elsewhere \cite{ZakeriSS09}. Prior to the SPEELS measurements, the structural, chemical and magnetic properties were checked by means of low energy electron diffraction, Auger electron spectroscopy and magneto optical Kerr effect measurements. The Fe films reveal the expected structural and magnetic properties well-known from literature \cite{Elmers95, Sander97,Elmers00}.
The SPEELS measurements were performed using our high performance spectrometer with an energy resolution well below 20 meV and a beam polarization of about 70$\pm10\%$ \cite{Ibach03}.

\begin{figure}[h]
    \vspace{12pt} \center
    \resizebox*{0.95\columnwidth}{!}{\includegraphics{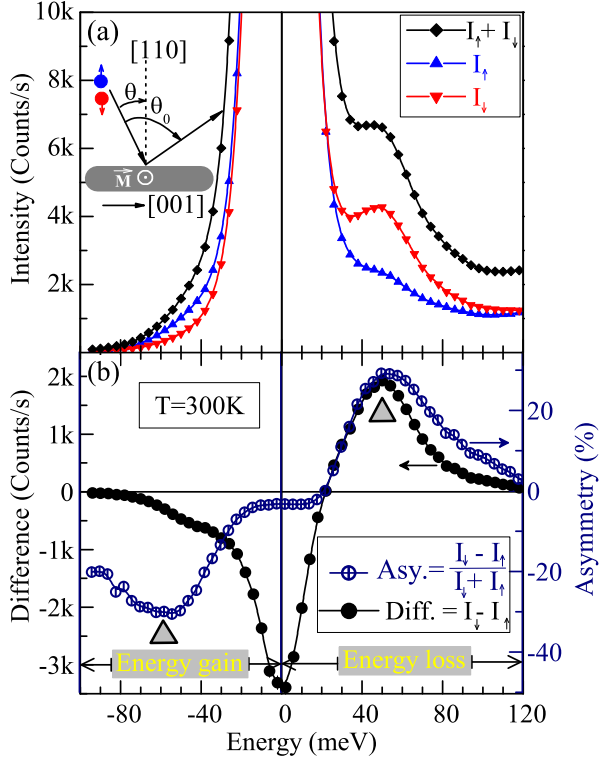}}
    \caption{\label{Fig1}(color online). (a) SPEEL-spectra measured on a 2 ML Fe film  epitaxially grown on W(110). The incoming electron beam had an energy of E$_0$=4 eV.  Inset shows the geometry of our SPEELS experiment. The spin-polarized electron beam is scattered along the [001]-direction of the Fe(110) surface in the magnetic remanent state. The scattering angle is kept at $\theta_0 = 80^{\circ}$. By changing the incident angle $\theta$, the in-plane wave vector transfer
parallel to the surface, $\Delta K^{\parallel}$, can be adjusted ($\Delta K^{\parallel}=k_f \sin(\theta_0-\theta)-k_i \sin(\theta)$, where $k_i$ and $k_f$ are the initial and final momentums of the electrons, respectively). For this experiment it was chosen to be $\Delta K^{\parallel}$=0.5 \AA$^{-1}$.
(b) The difference and  asymmetry spectra. The big triangles show the peak position due to the SW creation and annihilation taking place in energy loss and gain, respectively.}
\end{figure}

\begin{figure}[b]
    \vspace{12pt} \center
    \resizebox*{0.95\columnwidth}{!}{\includegraphics{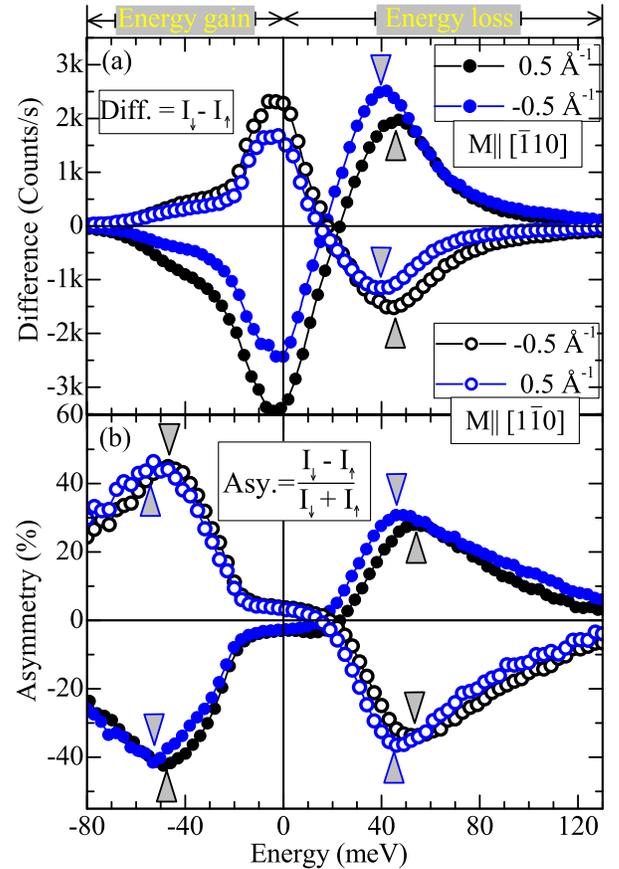}}
    \caption{\label{Fig2}(color online). (a) Series of difference, Diff.= I$_{\downarrow}$-I$_{\uparrow}$, and (b) asymmetry, Asy.=$\frac{I_{\downarrow}-I_{\uparrow}}{I_{\downarrow}+I_{\uparrow}}$, SPEEL spectra measured for $\Delta K^{\parallel}$=$\pm$0.5 \AA$^{-1}$ on a 2 ML Fe at RT. The filled symbols are for $\vec{M}\parallel[\bar{1}10]$ and the open ones are for $\vec{M}\parallel [1\bar{1}0]$. The spectra are recorded at a beam energy of E$_0$=4.163 eV with an energy resolution of 19 meV. The big triangles mark the peak positions of SW creations and annihilations, taking place at energy loss and gain, respectively.}
\end{figure}

A schematic representation of the scattering geometry is given in the inset of Fig. 1(a).
A well-defined monochromatized spin-polarized electron beam is scattered from the sample and the electron energy loss and gain spectra
are measured as a function of the in-plane momentum transfer ($\Delta K^{\parallel}$) for both spin orientations of the incoming electron beam (up$\uparrow$ and down$\downarrow$).
The surface SWs are excited in a virtual spin-flip scattering process
\cite{Plihal99,Vollmer03,Tang07,Prokop}.
The conservation of the angular momentum during the scattering prohibits SW excitations for incoming electrons with a spin polarization antiparallel to the sample magnetization ($I_{\uparrow}$). Hence, only electrons having
minority spin character ($I_{\downarrow}$) can create SWs. The  electrons with majority spin character ($I_{\uparrow}$) can, in principle, annihilate the thermally excited SWs while gaining energy. These facts lead to a peak in the minority spin channel in the energy loss region and a peak in the majority spin channel in the energy gain region (this is in analogy to the Stokes and ani-Stokes peaks in a Raman/Brillouin light scattering experiment). Figure 1(a) shows a typical SPEEL spectra measured at $\Delta K^{\parallel}$=0.5 \AA$^{-1}$ on a 2 ML Fe film. The amplitude of the peak due to the SW annihilation (in the energy gain region) is much smaller than the one caused by the SW creation. This is due to the fact that the probability of having thermally excited SWs in the system is given by the Boltzmann factor, which is about 0.01--0.1 at RT. This gives rise to a large peak in the energy loss region and a small dip in the energy gain region of the difference spectra.  However, both features can be seen clearly in the asymmetry curves, where the $Asy.=\frac{I_{\downarrow}-I_{\uparrow}}{I_{\downarrow}+I_{\uparrow}}$ is plotted as a function of energy for both loss and gain regions. In Fig. 1(b), the difference and asymmetry curves are presented. The big triangles mark the peak positions due to the spin-wave creation and annihilation processes.

In the absence of the DM interaction the spin-wave dispersion has to be symmetric with respect to the energy axis, meaning that measuring the SW spectra for negative wave vector transfers has to result in the same excitation energy as the one measured at positive wave vector transfers; $\Delta$E=E($\Delta K^{\parallel}$)-E($-\Delta K^{\parallel})$=0.

Figure 2 shows a series of difference and asymmetry curves measured on a 2 ML Fe film on W(110) at RT. The full symbols are the results of measurements when the magnetization is pointing along the [$\bar{1}$10]-direction. One clearly sees that for $\Delta K^{\parallel}$=0.5 \AA$^{-1}$ the SW creation peak (energy loss) is at higher energies, whereas the SW annihilation peak (energy gain) is at lower ones (it can be seen better in the asymmetry curves). The situation is totally reversed for negative wave-vector transfers i.e. $\Delta K^{\parallel}$=-0.5 \AA$^{-1}$; the SW annihilation peak is at higher energies and SW creation peak is at lower energies now. If this effect is caused by an uncertainty in the wave vector transfer, due to the stray fields induced bending of the electron beam in two different experiments, one would expect the same effect in the gain and loss regions (increase or decrease in both energies).
 The reversed phenomena in energy gain and loss regions indicate that this effect cannot be due to a slightly different electron trajectory in two different experiments. \\
Another argument, which clarifies that this is an intrinsic property of the system comes from measuring the same spectra for opposite magnetization directions. In magnetism, reversing the sample magnetization is equivalent to time inversion (in our experiment it basically means that the beam source and the detector are interchanged). The data for magnetization along the [1$\bar{1}$0]-direction are shown by open symbols in Fig . 2. In the case of reversed magnetization the SW excitation peak for $\Delta K^{\parallel}$=-0.5 \AA$^{-1}$ is at higher energies with respect to the one for $\Delta K^{\parallel}$=0.5 \AA$^{-1}$. This clearly indicates that having a slightly different energy for the SWs propagating along the [001]-direction with respect to the ones propagating along the [00$\bar{1}$]-direction is an intrinsic property of the SWs in this particular system. Based on the spin wave theory the symmetric exchange interaction cannot lead to any degeneracy breaking of the spin waves. One may think about presence of the dipolar interaction that is responsible for the unidirectional Damon-Eshbach surface modes \cite{Grunberg}. In this case the energy difference should be about 0.1 meV, which is much smaller than values observed in our experiment. Finally, we conclude that the presence of DM interaction breaks the degeneracy of the SW energies and leads to different energies for a given $\Delta K_{\parallel}$. Therefore, the assumption $\Delta E(\Delta K_{\parallel})= \Delta E(-\Delta K_{\parallel})$ is not valid here anymore.

It is worth to mention that measurements on a 20 ML thick sample showed also an energy asymmetry. The values of the energy asymmetry in this case are smaller than the ones measured for the double-layer. This observation reveals two facts: (i) since SPEELS is only sensitive to the top most layer(s), this effect is more likely a surface effect and is preserved up to even 20 ML thick films, (ii) this effect has nothing to do with the stray fields caused by the sample, because the stray fields strength is proportional the film thickness. If this effect was caused by stray fields, one would expect a larger effect for the thicker films.

The energy asymmetry, $\Delta E=E(\Delta K^{\parallel})-E(-\Delta K^{\parallel})$, induced by DM interaction varies with the in-plane wave vector transfer. In Fig. 3(a) the energy asymmetry is plotted as a function of the in-plane wave vector transfer, $\Delta K_{\parallel}$. Our data show that $\Delta E$ has a distinct maximum at $\pm0.5<\Delta K^{\parallel}<\pm1$.

\begin{figure}[t]
    \vspace{12pt} \center
    \resizebox*{0.95\columnwidth}{!}{\includegraphics{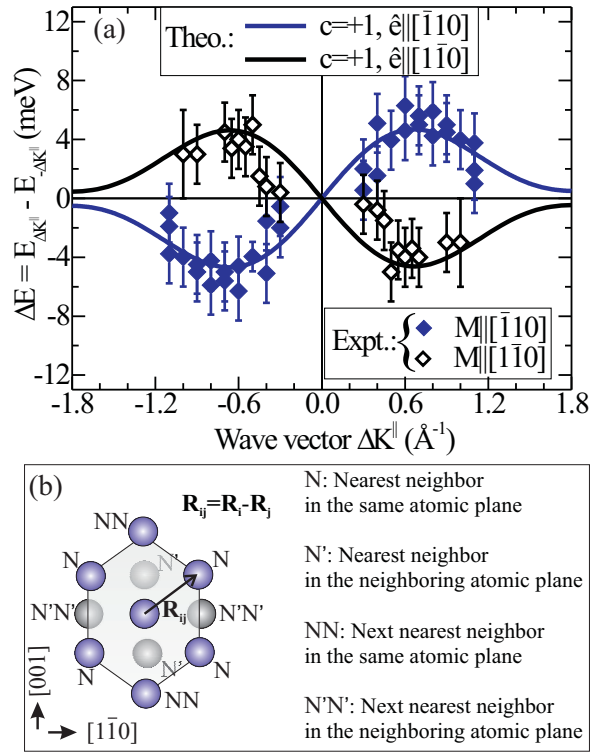}}
    \caption{\label{Fig3}(color online)
(a) The energy asymmetry as a function of wave vector transfer. The symbols are the measured values for two different magnetization directions and the solid curves are the fits using Eq. (2) for $c=+1$ and different magnetization directions. The error bars represent both the statistical and systematic uncertainties. (b) Real space representation of a 2 ML Fe slab.}
\end{figure}

Now, we attempt to estimate the amplitude of the DM vectors from our experimental data. By starting with a simple classical description of the SWs, the modified Heisenberg Hamiltonian in the presence of the DM interaction can be written as: $H=\sum_{i\neq j}J_{ij}\vec{S_i}\cdot \vec{S_j}-K_{eff}\sum_{i} (\vec{S_i}\cdot \hat{e})^2+\sum_{i\neq j}\vec{D_{ij}}\cdot \vec{S_i}\times \vec{S_j}$. Here the first term is the symmetric exchange interaction ($J_{ij}$ is the exchange coupling constant between spins $\vec{S_i}$ and $\vec{S_j}$), the second term is the magnetic anisotropy energy (MAE) term ($K_{eff}$ is the effective MAE constant with an easy axis along $\hat{e}$) and the last term is the DM interaction term ($D_{ij}$ are the DM vectors). The last term is the only one, which leads to an asymmetric dispersion relation. Using the same notation as in Ref. \cite{Udvardi09}, the asymmetry in the SW energies, $\Delta E=E_{DM}(\textbf{q})-E_{DM}(-\textbf{q})$, reads:

\begin{eqnarray}
	\Delta E&=&2c \sin^2\theta\sum_{i\neq j}\left(\vec{D}_{ij} \cdot\hat{e}\right) \sin\left[\textbf{q}\cdot(\vec{R}_i-\vec{R}_j)\right],
\end{eqnarray}

where \textbf{q} is the wave vector of the SWs (in our case SWs are propagating along the [001]-direction, therefore $\left|\textbf{q}\right|=\Delta K^{\parallel}$), c=$\pm$1 is the chirality rotation index (being +1 for right rotating sense and -1 for the left rotating one), $\hat{e}$ is the unit-vector of the magnetization $\vec{M}$, $\theta$ is the relative angle between moments and $\hat{e}$  and $\vec{R}_i (\vec{R}_j)$ is the position vector of site $i(j)$. For a double-layer slab Eq. (1) can be simplified to:

\begin{eqnarray}
	\Delta E=\pm4c\left[\left(2D^x_{1}+\acute{D^x_{1}}\right) \sin\left(\frac{\Delta K^{\parallel}a}{2}\right)\right. \nonumber\\ \left.+D^x_{2}\sin\left(\Delta K^{\parallel} a\right)\right].
\end{eqnarray}

Here the $\pm$ sign stands for different magnetization directions, $a$ is the inter-atomic-distance being 3.16 \AA~and $D^x_i=\sin^2 \theta \vec{D}_i\cdot \hat{e}$ ($\acute{D}^x_{i}=\sin^2 \theta \acute{\vec{D}}_i\cdot \hat{e}$) is the longitudinal component of the DM vector of the $i^{th}$ neighbors in the same atomic plane (in the neighboring atomic plane, see Fig. 3(b)). The maximum $|\Delta E|$ observed in our experiments, taking place at $\pm0.5<\Delta K^{\parallel}<\pm1$, is in line with the fact that Eq. (2) has also a local extremum at $\pm\pi/2a\approx\pm0.5<\Delta K^{\parallel}<\pm\pi/a\approx\pm1$. By fitting the experimental data with Eq. (2) for different magnetization directions ($\hat{e}\parallel[1\bar{1}0]-$ and $\hat{e}\parallel[\bar{1}10]-$ direction) one finds $|2D^x_1+\acute{D^x_{1}}|$=0.9(3) meV and $|D^x_2|$=0.5(3) meV. Note that the fits are performed for $c=+1$ meaning that the spin spiral structure, which is formed by DM interaction has right rotating sense, in agreement with the recent experimental results good obtained by spin-polarized scanning tunneling microscopy \cite{Meckler}.

This is the first direct experimental determination of the DM vector components on each individual atomic site. The measured values are smaller than the theoretically predicted values for an Fe monolayer \cite{Udvardi09}.  However, those calculations were done for the monolayer Fe without considering the temperature effects, whereas our experimental results are obtained for the double-layer at RT by employing a simple model. This discrepancy might also be due to the fact that in the calculations the electron-magnon and phonon-magnon coupling are not considered.

In summary, we showed that the DM interaction lifts the degeneracy of the SWs and leads to an asymmetric spin-wave dispersion relation. The DM interaction induced energy asymmetry of SW energies is measured for an Fe double-layer grown on W(110) using SPEELS. It is shown that the DM interaction is preserved even up to RT. An extended Heisenberg spin Hamiltonian is used to obtain the component of the DM vectors from the experimentally measured energy asymmetry. Our results, which reveal the importance of the antisymmetric exchange interaction, provide a new insight into the spin-dynamics in magnetic nanostructures and would contribute to a better understanding of magnetism on the nano scale.

The authors appreciate the fruitful discussions with D. L. Mills and L. Szunyogh.

\end{document}